\documentclass[aps,pra,preprintnumbers,showpacs,tightenlines]{revtex4}

\usepackage{amssymb}
\usepackage{amsmath}
\usepackage{graphicx}
\usepackage{epsfig}
\usepackage{subfigure}
\usepackage{amsfonts}
\usepackage{CJK}

\begin{document}

\title{Preparing Greenberger-Horne-Zeilinger Entangled Photon
Fock States of Three Cavities Coupled by a Superconducting Flux Qutrit}

\author{Zhen-Fei Zheng$^{2}$, Qi-Ping Su$^{1}$, and Chui-Ping Yang$^{1,3}$\thanks{E-mail: yangcp@hznu.edu.cn}}
\thanks{E-mail: yangcp@hznu.edu.cn}

\address{$^1$Department of Physics, Hangzhou Normal University,
Hangzhou, Zhejiang 310036, China}

\address{$^2$Jing Hengyi Honors College, Hangzhou Normal University, Hangzhou, Zhejiang 310036, China}

\address{$^3$State Key Laboratory of Precision Spectroscopy, Department of Physics,
East China Normal University, Shanghai 200062, China}

\date{\today}

\begin{abstract}
We propose a way to prepare Greenberger-Horne-Zeilinger (GHZ) entangled photon Fock states
of three cavities, by using a superconducting flux qutrit coupled to the cavities. This
proposal does not require the use of classical microwave pulses and
measurement during the entire operation. Thus, the operation is greatly
simplified and the circuit engineering complexity and cost is much reduced. The proposal is
quite general and can be applied to generate three-cavity GHZ entangled photon Fock states
when the three cavities are coupled by a different three-level physical system
such as a superconducting charge qutrit, a transmon qutrit, or a quantum dot.
\end{abstract}

\pacs{03.67.-a, 42.50.Pq, 85.25.-j} \maketitle
\date{\today}

\begin{center}
\textbf{I. INTRODUCTION}
\end{center}

Entanglement is one of the important properties, through which a quantum
physical system can be identified to be distinguished from a classical
physical system. On the other hand, entanglement is one of the cornerstones
in building up a quantum machine. It plays an important role in both quantum
information processing (QIP) and quantum communication (e.g., quantum
teleportation, quantum secret sharing, quantum key distribution and so on).
Over the past decade, experimental preparation of entanglement has been
reported with various of physical systems, such as eight photons via linear
optical devices [1], fourteen trapped ions [2], three spins [3], two atoms
in microwave cavity QED [4], two atoms plus one cavity mode [5], and two
excitions in a single quantum dot [6]. In addition, many schemes have been
proposed for generating entangled states of superconducting qubits
based on cavity QED [7] or via capacitive couplings [8].
Moreover, various two-qubit or three-qubit entangled states have been experimentally
demonstrated with superconducting qubits coupled to a single cavity [9-13].

Superconducting devices [14-16] play significant roles in scalable
quantum computing. The physical system composed of cavities and
superconducting qubits is one of the most promising candidates for QIP [14,15].
Superconducting qubits, such as flux, charge and phase qubits, have long
decoherence time [17,18] and experiments have realized quantum operations in
single and multiple superconducting qubits with states read out [19-23]. A
cavity (such as coplanar waveguide, microstrip resonator and lumped-circuit
resonator and so on) can act as a quantum bus, which can mediate
long-distance and fast interaction between distant superconducting qubits [24-29].
In addition, superconducting qubits and microwave resonators can be
fabricated with modern technology. Furthermore, the strong coupling limit
between the superconducting qubits and the cavity field was predicated
earlier [29,30] and has been experimentally demonstrated [31,32].

So far, many theoretical proposals have been presented for the preparation
of Fock states, coherent states, squeezed states, the Sch\"ordinger Cat
state, and an arbitrary superposition of Fock states of a single
superconducting cavity [33-36]. Recently, there is much interest in
generation of entangled states of qubits or photons in multiple cavities
because of their importance in realizing large-scale QIP within cavity QED.
Several theoretical proposals for generation of entangled photon Fock states
of two resonators by a superconducting coupler qubit [37,38] have been
presented. A theoretical proposal for the manipulation and generation of
nonclassical microwave field states as well as the creation of controlled
multipartite entanglement with two resonators coupled by a superconducting
qubit has been also presented [39]. Based on the resonant interaction with a
tunable superconducting phase qutrit coupled to two resonators, an efficient
proposal for generating NOON states has been proposed recently [40].
Moreover, by using a phase qutrit coupled to two resonators, a recent
experimental demonstration of an entangled NOON state of photons in two
superconducting microwave resonators has been reported [41]. However, we
note that how to prepare entangled states of photons in more than two
cavities has not been thoroughly investigated.

In this paper, we propose a way for preparing Greenberger-Horne-Zeilinger
(GHZ) entangled photon Fock states of three cavities coupled to a
superconducting flux qutrit. As shown below, this proposal has the following
advantages: (i) No classical microwave pulse is needed during the
entanglement preparation; (ii) Neither measurement on the states of the
coupler qutrit nor detection on the states of photons in cavities is
required; and (iii) entangled photon Fock states of three cavities can be
generated using one coupler qutrit and two steps of operation only.

To the best of our knowledge, our proposal is the first one to demonstrate
that a GHZ entangled photon Fock state for three cavities can be prepared by
using a superconducting flux qutrit [24,29,42-44] as a coupler and without need of both
classical microwave pulses and measurement. Since neither classical
microwave pulse nor measurement is needed and only one coupler qutrit is
used, the operation is greatly simplified, and the circuit engineering
complexity and cost is much reduced, by using the present proposal.
Furthermore, the method presented here is quite general, and can be applied
to generate three-cavity GHZ entangled photon Fock states when the three
cavities are coupled by a different three-level physical system such as a
superconducting charge qutrit, a transmon qutrit, or a quantum dot.

This paper is organized as follows. In Sec. II, we introduce the Raman
resonant coupling induced due to a coupler flux qutrit interacting with two
cavities. In Sec. III, we discuss how to generate the GHZ entangled photon
Fock state of three cavities and then give a brief discussion of the
experimental issues. In Sec. IV, we give a discussion of the fidelity and
possible experimental implementation. A concluding summary is given in Sec.
V.

\begin{figure}[tbp]
\begin{center}
\includegraphics[bb=67 486 532 718, width=12.5 cm, clip]{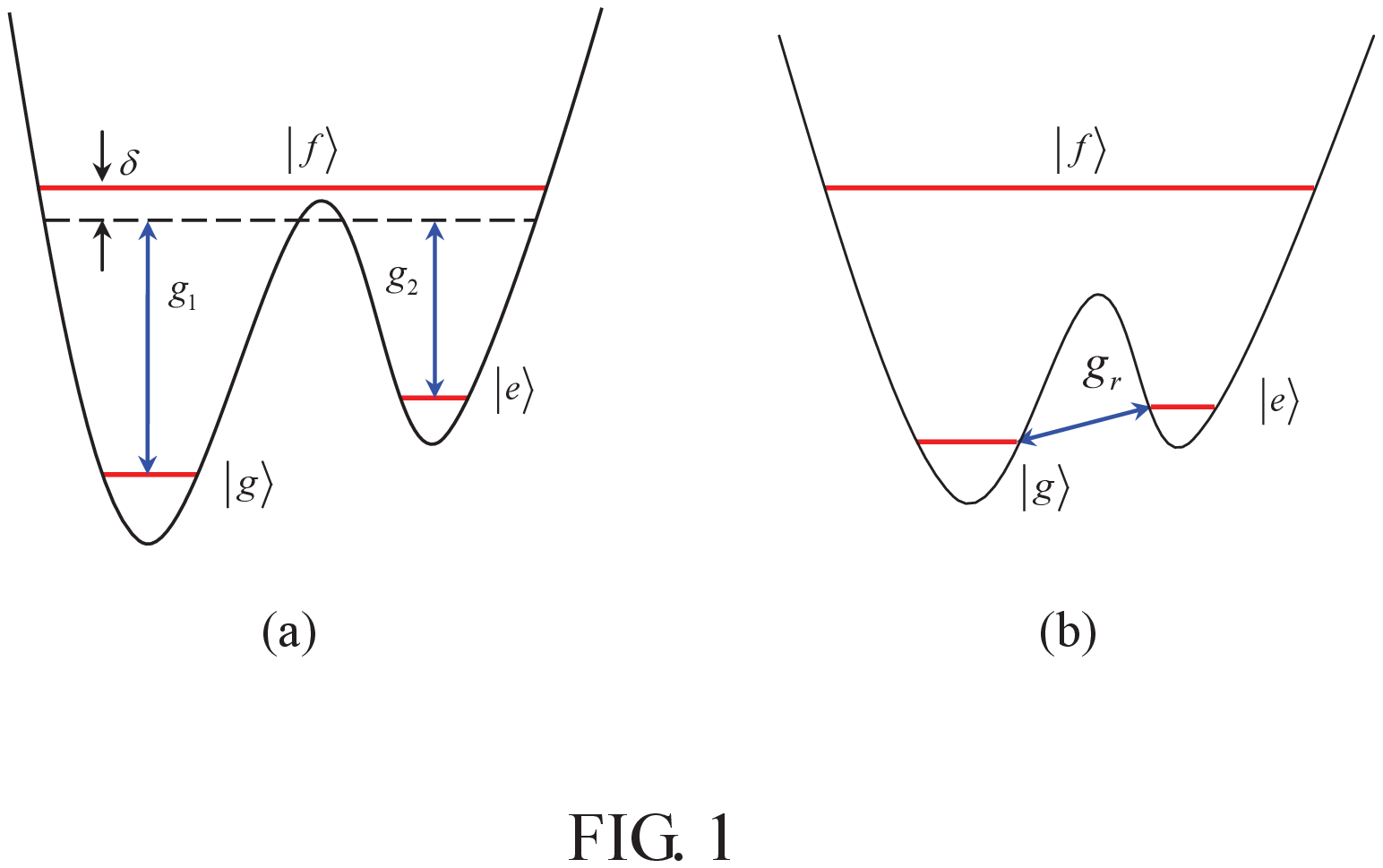} %
\vspace*{-0.08in}
\end{center}
\caption{(Color online) (a) Illustration of cavity $1$ dispersively coupled
to the $\left| g\right\rangle \leftrightarrow \left| f\right\rangle $
transition and cavity 2 dispersively coupled to the $\left| e\right\rangle
\leftrightarrow \left| f\right\rangle $ transition of qutrit $A$ (a $\Lambda$%
-type three-level flux qutrit). The transition between the two lowest levels
can be made weak by increasing the barrier between two potential wells. (b)
Illustration of cavity $3$ resonantly coupled to the transition between the
two lowest levels of qutrit $A$. The resonant coupling constant $g_r$ can be
increased by increasing the dipole coupling matrix element between the two
lowest levels, which can be reached by lowering the potential barrier [46].}
\label{fig:1}
\end{figure}

\begin{center}
\textbf{II. RAMAN RESONANT COUPLING}
\end{center}

Consider a system which contains a superconducting flux qutrit $A$ and two
cavities. Cavity $1$ is coupled to the $\left| g\right\rangle
\leftrightarrow \left| f\right\rangle $ transition with a coupling constant $%
g_1$ and a detuning $\delta =\omega _{fg}-\omega _{c1}$ and cavity $2$ is
coupled to the $\left| e\right\rangle \leftrightarrow \left| f\right\rangle $
transition with a coupling constant $g_2$ and a detuning $\delta =\omega
_{fe}-\omega _{c2}$ [Fig.~1(a)]. Here, $\omega _{c1}$ ($\omega _{c2}$) is
the frequency of cavity $1$ ($2$) while $\omega _{fg}$ ($\omega _{fe}$) is
the $\left| g\right\rangle \leftrightarrow \left| f\right\rangle $ ($\left|
e\right\rangle \leftrightarrow \left| f\right\rangle $) transition frequency
of qutrit $A$. In addition, assume that each cavity is highly detuned
(decoupled) from the transition between any other two levels of the qutrit $%
A.$ In the interaction picture, the Hamiltonian of the whole system is given
by
\begin{equation}
H=g_1(e^{i\delta t}a_1S_{fg}^{+}+h.c)+g_2(e^{i\delta t}a_2S_{fe}^{+}+h.c.),
\end{equation}
where $S_{fg}^{+}=\left| f\right\rangle \left\langle g\right| $, $%
S_{fe}^{+}=\left| f\right\rangle \left\langle e\right| ,$ and $a_1$ ($a_2$)
is the photon annihilation operator for cavity $1$ ($2$).

Suppose that cavity $1$ is dispersively coupled to the $\left|
g\right\rangle \leftrightarrow \left| f\right\rangle $ transition (i.e., $%
\delta \gg g_1$) and cavity $2$ is dispersively coupled to the $\left|
e\right\rangle \leftrightarrow \left| f\right\rangle $ transition (i.e., $%
\delta \gg g_2$). Under this condition, the Hamiltonian (1) reduces to
\begin{equation}
H_{eff}=H_0+H_I
\end{equation}
with
\begin{equation}
H_0=-\frac{g_1^2}\delta a_1^{+}a_1\left| g\right\rangle \left\langle
g\right| -\frac{g_2^2}\delta a_2^{+}a_2\left| e\right\rangle \left\langle
e\right| ,
\end{equation}
\begin{equation}
H_I=-\frac{g_1g_2}\delta (a_1^{+}a_2S_{eg}^{-}+h.c.),
\end{equation}
where $S_{eg}^{-}=\left| g\right\rangle \left\langle e\right| .$ The two
terms in Eq. (3) are ac-stark shifts of the energy levels $\left|
g\right\rangle $ and $\left| e\right\rangle $ which are induced by the mode
of cavity $1$ and the mode of cavity $2$, respectively; while Eq. (4)
describes the Raman resonant $\left| g\right\rangle \leftrightarrow \left|
e\right\rangle $ coupling caused due to the two-cavity cooperation.

Performing an unitary transform $U=e^{-iH_0t},$ we obtain

\begin{eqnarray}
\widetilde{H}_I\!\!\!\!&=&\!\!\!\!U^{+}H_IU  \nonumber \\
\!\!\!\! &=&\!\!\!\!-\frac{g_1g_2}\delta (e^{-i\frac{g_1^2}\delta
a_1^{+}a_1t}a_1^{+}a_2S_{eg}^{-}e^{i\frac{g_2^2}\delta a_2^{+}a_2t}+h.c.).
\end{eqnarray}

Denote $\left| 0\right\rangle _j$ and $\left| 1\right\rangle _j$ as the
vacuum state and the single-photon state for cavity $j$ ($j=1,2$),
respectively. One can see that under the Hamiltonian (5), the state $\left|
e\right\rangle \left| 0\right\rangle _1\left| 1\right\rangle _2$ evolves
within a Hilbert space formed by the two orthogonal states $\left|
e\right\rangle \left| 0\right\rangle _1\left| 1\right\rangle _2$ and $\left|
g\right\rangle \left| 1\right\rangle _1\left| 0\right\rangle _2$. In this
Hilbert space, the Hamiltonian (5) can be expanded as

\begin{equation}
\widetilde{H}_I=\left(
\begin{array}{cc}
0 & -e^{i\varphi t}g_1g_2/\delta \\
-e^{-i\varphi t}g_1g_2/\delta & 0
\end{array}
\right) ,
\end{equation}
where $\varphi =g_1^2/\delta -g_2^2/\delta .$ For $g_1=g_2=g$ (achievable by
tuning the coupling capacitance $C_1$ between qutrit $A$ and cavity $1$ as
well as the coupling capacitance $C_2$ between qutrit $A$ and cavity $2,$
see Fig. 2 below)$,$ the matrix (6) becomes

\begin{equation}
\widetilde{H}_I=\left(
\begin{array}{cc}
0 & -g^2/\delta \\
-g^2/\delta & 0
\end{array}
\right) .
\end{equation}
One can easily find that under the Hamiltonian (7), the time evolution of
the state $\left| e\right\rangle \left| 0\right\rangle _1\left|
1\right\rangle _2$ is described by

\begin{equation}
\left| e\right\rangle \left| 0\right\rangle _1\left| 1\right\rangle
_2\rightarrow \cos (\frac{g^2}\delta t)\left| e\right\rangle \left|
0\right\rangle _1\left| 1\right\rangle _2+i\sin (\frac{g^2}\delta t)\left|
g\right\rangle \left| 1\right\rangle _1\left| 0\right\rangle _2.
\end{equation}
We now return to the original interaction picture by applying an unitary
transformation $e^{-iH_0t}$ to the right side of Eq. (8). It can be found
that in the original interaction picture, the time evolution of the state $%
\left| e\right\rangle \left| 0\right\rangle _1\left| 1\right\rangle _2$
remains the same as Eq.~(8), except for a common phase $e^{ig^2t/\delta }$
which appears on both terms $\left| e\right\rangle \left| 0\right\rangle
_1\left| 1\right\rangle _2$ and $\left| g\right\rangle \left| 1\right\rangle
_1\left| 0\right\rangle _2$ of Eq.~(8).

The result (8) obtained here will be employed for the preparation of the
three-cavity GHZ entangled photon Fock states, as discussed below.

\begin{center}
\textbf{III. PREPARING THREE-CAVITY GHZ-TYPE ENTANGLED PHOTON FOCK STATES}
\end{center}

\begin{figure}[tbp]
\begin{center}
\includegraphics[bb=214 465 428 655, width=8.5 cm, clip]{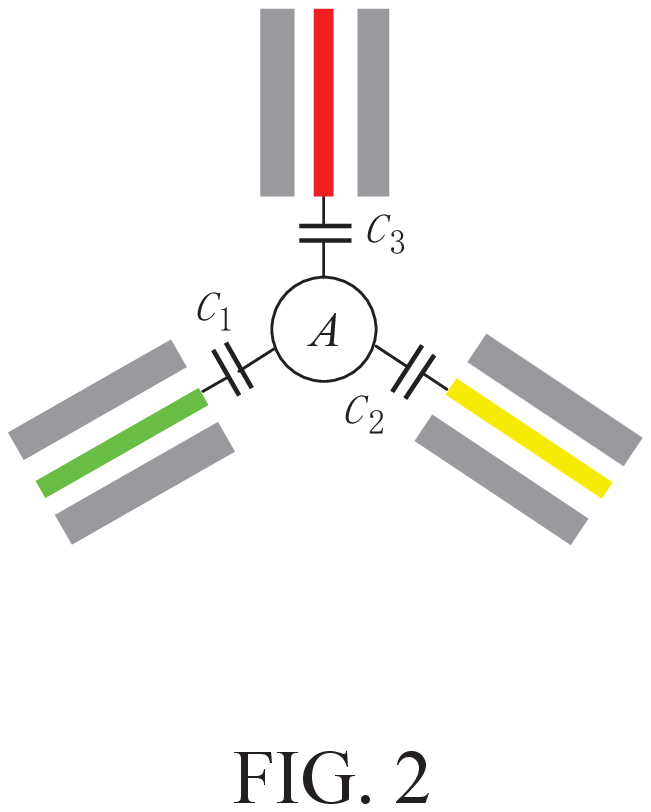} %
\vspace*{-0.08in}
\end{center}
\caption{(Color online) Setup for triple cavities coupled by a
superconducting flux qutrit. Each cavity here is a one-dimensional coplanar
waveguide transmission line resonator. The circle A represents a
superconducting flux qutrit, which is capacitively coupled to cavity $j$ via
a capacitance $C_j$ ($j=1,2,3$).}
\label{fig:2}
\end{figure}

Consider a superconducting flux qutrit $A$ coupled to three cavities, which
is illustrated in Fig.~2. Initially, qutrit $A$ is decoupled from each
cavity, which can be reached by a prior adjustment of the level spacings of
qutrit $A$, such that each cavity is highly detuned from the transition
between any two levels of qutrit $A$. Note that for a superconducting
qutrit, the level spacings can be rapidly adjusted by varying external
control parameters (e.g., magnetic flux applied to phase, transmon, or flux
qutrits, see e.g. Refs. [17], [44], and [45]). In addition, assume that qutrit $A$ is
initially in the state $\left| e\right\rangle $; and cavities $1$ and $3$
are initially in the vacuum state while cavity $2$ is initially in a
single-photon state.

The procedure for generating the GHZ entangled photon Fock state of triple
cavities is listed as follows.

Step (i): Adjust the level spacing of qutrit $A$ such that cavity $3$ is
decoupled from qutrit $A$ but cavity $1$ is dispersively coupled to the $%
\left| g\right\rangle \leftrightarrow \left| f\right\rangle $ transition and
cavity $2$ is dispersively coupled to the $\left| e\right\rangle
\leftrightarrow \left| f\right\rangle $ transition, as shown in Fig.~1(a).
When the conditions mentioned in Sec. II are met, the time evolution of the
state $\left| e\right\rangle \left| 0\right\rangle _1\left| 1\right\rangle _2
$ is given by Eq.~(8), from which one can see that if qutrit $A$ interacts
with cavities $1$ and $2$ for a given time $t_1=\delta \pi /\left(
4g^2\right) $, then the initial state $\left| e\right\rangle \left|
0\right\rangle _1\left| 1\right\rangle _2\left| 0\right\rangle _3$ of the
whole system evolves into

\begin{equation}
\frac 1{\sqrt{2}}\left( \left| e\right\rangle \left| 0\right\rangle _1\left|
1\right\rangle _2+i\left| g\right\rangle \left| 1\right\rangle _1\left|
0\right\rangle _2\right) \left| 0\right\rangle _3.
\end{equation}

Step (ii): Adjust the level spacing of qutrit $A$ such that cavities $1$ and
$2$ are now decoupled from qutrit $A$, but cavity $3$ is resonant with the
transition between the two lowest levels $\left| g\right\rangle $ and $%
\left| e\right\rangle $ of qutrit $A$ [Fig.~1(b)], with a resonant coupling
constant $g_r$ (which can be increased by rapidly lowering the potential
barrier between the two potential wells [46]). One can see that if qutrit $A$
interacts with cavity $3$ for a given time $t_2=\pi /\left( 2g_r\right) ,$
the state $\left| g\right\rangle \left| 0\right\rangle _3$ remains unchanged
while the state $\left| e\right\rangle \left| 0\right\rangle _3$ changes to $%
-i\left| g\right\rangle \left| 1\right\rangle _3$ in the interaction picture
[47]. As a result, the state (9) becomes
\begin{equation}
\left( -i\right) \frac 1{\sqrt{2}}\left( \left| 0\right\rangle _1\left|
1\right\rangle _2\left| 1\right\rangle _3-\left| 1\right\rangle _1\left|
0\right\rangle _2\left| 0\right\rangle _3\right) \left| g\right\rangle .
\end{equation}

The result (10) demonstrates that a three-cavity GHZ entangled photon Fock
state is prepared while qutrit $A$ is disentangled from each cavity after
the above operations. Note that after the operation of step (ii), the level
spacings of qutrit $A$ need to be adjusted back to the original
configuration (the one before the entanglement preparation), such that
qutrit $A$ is decoupled from the cavities. Thus, after the operation, the
prepared GHZ state is maintained because of no qutrit-cavity interaction.

For the method to work, the following conditions need to be satisfied:

(i) The occupation probability $p$ of the level $\left| f\right\rangle $ for
the qutrit during step (i) is given by $p\simeq 4/\left[ 4+\left( \delta
/g\right) ^2\right] $ [48], which needs to be negligibly small in order to
reduce the operation error.

(ii) The total operation time $\tau $, given by
\begin{eqnarray}
\tau =\delta \pi /\left( 4g^2\right) +\pi /\left( 2g_r\right) +3t_d+t_b
\end{eqnarray}
(where $t_d$ is the typical time required for adjusting the qutrit level
spacings while $t_b$ is the typical time required for adjusting the
potential barrier), needs to be much shorter than the energy relaxation time
$T_1$ and dephasing time $T_2$ of the levels $\left| e\right\rangle $ (note
that the level $\left| f\right\rangle $ is almost not occupied during the
operation), such that decoherence, caused due to energy relaxation and
dephasing process of the qutrit, is negligible during the operation.

(iii) For cavity $i$ ($i=1,2,3$), the lifetime of the cavity mode is given
by $T_{cav}^i=\left( Q_i/\omega _{ci}\right) /\overline{n}_i,$ where $Q_i$
and $\overline{n}_i$ are the (loaded) quality factor and the average photon
number of cavity $i$, respectively. For the three cavities here, the
lifetime of the cavity modes is given by
\begin{equation}
T_{cav}=\frac 13\min \{T_{cav}^1,T_{cav}^2,T_{cav}^3\},
\end{equation}
which should be much longer than $\tau ,$ such that the effect of cavity
decay is negligible during the operation.

(iii) There exists an intercavity crosstalk coupling between any two
cavities, e.g., cavities $k$ and $l$, during each step of the operation,
which however can be made negligible as long as the cavity-cavity frequency
detuning $\Delta _{kl}$ is much larger than the intercavity crosstalk
coupling constant $g_{kl}$ between cavities $k$ and $l$ ($kl=12,13,23$).

Before ending this section, we should point out that adjusting the level
spacings of qutrit $A$ is unnecessary. Alternatively, the method can be
implemented with adjusting the frequencies of the cavities. It should be
noticed that rapid tuning of cavity frequencies has been demonstrated
recently in superconducting microwave cavities (e.g., in less than a few
nanoseconds for a superconducting transmission line resonator [49]).

\begin{center}
\textbf{IV. FIDELITY}
\end{center}

The proposed protocol for creating the state (10) involves two basic steps
of operation as discussed above.

(i) During the operation of step (i), cavity $1$ is dispersively coupled to
the $\left| g\right\rangle \leftrightarrow \left| f\right\rangle $
transition and cavity $2$ is dispersively coupled to the $\left|
e\right\rangle \leftrightarrow \left| f\right\rangle $ transition, while
cavity $3$ requires to be decoupled from qutrit $A$. In the interaction
picture, the interaction Hamiltonian governing this operation is given by
\begin{eqnarray}
H_{I,1} &=&g_1(e^{i\delta t}a_1S_{fg}^{+}+h.c)+g_2(e^{i\delta
t}a_2S_{fe}^{+}+h.c.)  \nonumber \\
&+&\sum_{j=1,3}g_{jfe}\left( e^{i\Delta
_{jfe}t}a_jS_{fe}^{+}+h.c.\right)  \nonumber\\
&+&\sum_{j=2,3}g_{jfg}\left( e^{i\Delta
_{jfg}t}a_jS_{fg}^{+}+h.c.\right)   \nonumber \\
&+&\sum_{j=1,2,3}g_{jeg}\left( e^{i\Delta
_{jeg}t}a_jS_{eg}^{+}+h.c.\right)  \nonumber \\
&+&\sum_{kl=12,13,23}g_{kl}\left(
e^{i\Delta _{kl}t}a_ka_l^{+}+h.c.\right) ,
\end{eqnarray}
where the subscript $j$ represents cavity $j$, $S_{eg}^{+}=\left|
e\right\rangle \left\langle g\right| ,$ and $\Delta _{kl}=\omega
_{cl}-\omega _{ck}$ with $\omega _{cl}$ ($\omega _{ck}$) being the frequency
of cavity $l$ ($k$). Let us give some explanation on Eq.~(13). The two terms
in the first line describe the dispersive interaction between cavity $1$ and
the $\left| g\right\rangle \leftrightarrow \left| f\right\rangle $
transition as well as the dispersive interaction between cavity $2$ and the $%
\left| e\right\rangle \leftrightarrow \left| f\right\rangle $ transition,
which induce the Raman resonant coupling between the two lowest levels $%
\left| g\right\rangle \leftrightarrow \left| e\right\rangle $ transition of
qutrit $A,$ as discussed in Sec. II. The term in the second line
represents the unwanted off-resonant coupling between cavity $j$ ($j=1,3$)
and the $\left| e\right\rangle \leftrightarrow \left| f\right\rangle $
transition, with a coupling constant $g_{jfe}$ and detuning $\Delta
_{jfe}=\omega _{fe}-\omega _{cj}$. The term in the third line
indicates the unwanted off-resonant coupling between cavity $j$ ($j=2,3$)
and the $\left| g\right\rangle \leftrightarrow \left| f\right\rangle $
transition, with a coupling constant $g_{jfg}$ and detuning $\Delta
_{jfg}=\omega _{fg}-\omega _{cj}.$ The term in the fourth line
represents the unwanted off-resonant coupling between cavity $j$ ($j=1,2,3$)
and the $\left| g\right\rangle \leftrightarrow \left| e\right\rangle $
transition, with a coupling constant $g_{jeg}$ and detuning $\Delta
_{jeg}=\omega _{eg}-\omega _{cj}$ ($\omega _{eg}$ is the $\left|
g\right\rangle \leftrightarrow \left| e\right\rangle $ transition
frequency). The term in the last line represents the unwanted
intercavity crosstalk between any two of three cavities.

(ii) During the operation of step (ii), both cavities $1$ and $2$ require to
be decoupled from qutrit $A,$ while cavity $3$ is resonant with the $\left|
g\right\rangle \leftrightarrow \left| e\right\rangle $ transition. In the
interaction picture, the interaction Hamiltonian for this operation is given
by
\begin{eqnarray}
H_{I,2} &=&g_r\left( a_3S_{eg}^{+}+h.c.\right) +\sum_{j=1,2}g_{jeg}^{\prime
}\left( e^{i\Delta _{jeg}^{\prime }t}a_jS_{eg}^{+}+h.c.\right)   \nonumber \\
&&+\sum_{j=1,2,3}g_{jfg}^{\prime }\left( e^{i\Delta _{jfg}^{\prime
}t}a_jS_{fg}^{+}+h.c.\right) \nonumber \\
&&+\sum_{j=1,2,3}g_{jfe}^{\prime }\left(
e^{i\Delta _{jfe}^{\prime }t}a_jS_{fe}^{+}+h.c.\right) +\varepsilon ,
\end{eqnarray}
where the first term in the first line represents the resonant interaction
between cavity $3$ and the $\left| g\right\rangle \leftrightarrow \left|
e\right\rangle $ transition; all other terms are the unwanted interaction
between the qutrit and cavities; and $\varepsilon $ is the last term in
Eq.~(13) for the unwanted intercavity crosstalk. In addition, $%
g_{jkl}^{\prime }$ is the coupling constant between cavity $j$ and the $%
\left| l\right\rangle \leftrightarrow \left| k\right\rangle $ transition of
qutrit $A$ ($kl=fg,fe,eg$), and $\Delta _{jkl}^{\prime }=\omega
_{kl}^{\prime }-\omega _{cj}$ is the detuning between the frequency $\omega
_{cj}$ of cavity $j$ and the adjusted $\left| l\right\rangle \leftrightarrow
\left| k\right\rangle $ transition frequency $\omega _{kl}^{\prime }$.

When the dissipation and dephasing are included, the dynamics for the $k$th
step of operation is determined by

\begin{eqnarray}
\frac{d\rho }{dt} &=&-i\left[ H_{I,k},\rho \right] +\sum_{j=1}^3\kappa _j%
\mathcal{L}\left[ a_j\right] +\gamma _{k\varphi ,fe}\left( S_{fe}^z\rho
S_{fe}^z-\rho \right)
\nonumber \\
&&+\gamma _{kfe}\mathcal{L}\left[ S_{fe}^{-}\right]+\gamma _{k\varphi ,fg}\left( S_{fg}^z\rho S_{fg}^z-\rho
\right) +\gamma _{kfg}\mathcal{L}\left[ S_{fg}^{-}\right] \nonumber \\
&&+\gamma _{k\varphi
,eg}\left( S_{eg}^z\rho S_{eg}^z-\rho \right) +\gamma _{keg}\mathcal{L}%
\left[ S_{eg}^{-}\right] ,
\end{eqnarray}
where $\mathcal{L}\left[ \Lambda \right] =\Lambda \rho \Lambda ^{+}-\Lambda
^{+}\Lambda \rho /2-\rho \Lambda ^{+}\Lambda /2,$ with $\Lambda
=a_j,S_{fe}^{-},S_{fg}^{-},S_{eg}^{-}$; $S_{fe}^z=\left| f\right\rangle
\left\langle f\right| -\left| e\right\rangle \left\langle e\right| $, $%
S_{fg}^z=\left| f\right\rangle \left\langle f\right| -\left| g\right\rangle
\left\langle g\right| $, and $S_{eg}^z=\left| e\right\rangle \left\langle
e\right| -\left| g\right\rangle \left\langle g\right| $. In addition, $%
\kappa _j$ is the decay rate of the mode of cavity $j$ ($j=1,2,3$)$,$ $%
\gamma _{k\varphi ,fe}$ ($\gamma _{k\varphi ,fg}$) and $\gamma _{kfe}$ ($%
\gamma _{kfg}$) are the dephasing rate and the energy relaxation rate of the
level $\left| f\right\rangle $ of qutrit $A$ for the decay path $\left|
f\right\rangle \rightarrow \left| e\right\rangle $ ($\left| g\right\rangle $%
), respectively and $\gamma _{k\varphi ,eg}$ and $\gamma _{keg}$ are those
of the level $\left| e\right\rangle $ for the decay path $\left|
e\right\rangle \rightarrow \left| g\right\rangle $ (the superscript $k$ here
is used to indicate the dephasing rate and relaxation rate of qutrit $A$
during the $k$th step of operation).

The fidelity of the operation is given by
\begin{equation}
\mathcal{F}=\left\langle \psi _{id}\right| \widetilde{\rho }\left| \psi
_{id}\right\rangle ,
\end{equation}
where $\left| \psi _{id}\right\rangle $ is the state (10) of an ideal system
(i.e., without dissipation, dephasing, and crosstalk) and $\widetilde{\rho }
$ is the final density operator of the system when the operation is
performed in a realistic physical system.

We now numerically calculate the fidelity of the prepared GHZ entangled Fock
state. The parameters used in our numerical calculation for each step of
operation are listed below.

{\it Parameters for Step (i)---}For a superconducting flux qutrit with
three levels illustrated in Fig.~1, the typical transition frequency $\omega
_{fe}/2\pi $ or $\omega _{fg}/2\pi $ is between 4 and 20 GHz, and $\omega
_{eg}/2\pi $ can be made to be from $1$ GHz to $5$ GHz. Thus, for the flux
qutrit with three levels shown in Fig. 1(a) [corresponding to the operation
of step (i)], we choose $\omega _{eg}/2\pi \sim 5$ GHz and $\omega
_{fe}/2\pi \sim 10.0$ GHz. In addition, with appropriate design of the
qutrit system, one can have $\phi _{fg}\sim \phi _{fe}\sim 10\phi _{eg},$
where $\phi _{ij}$ represents the dipole coupling matrix element between the
two levels $\left| i\right\rangle $ and $\left| j\right\rangle $ with $ij\in
\{fg,fe,eg\}.$ As a result, we have $g_{1eg}\sim 0.1g_1,$ $g_{1fe}\sim g_1;$
$g_{2eg}\sim 0.1g_2,$ $g_{2fg}\sim g_2;$ and $g_{3eg}\sim 0.1g_r,$ $%
g_{3fe}\sim g_{3fg}\sim g_r$ [50]$.$ Here, $g_1=g_2=g$ (set previously).
Other parameters used in the numerical calculation for this step of
operation are: (i) $g_r/2\pi =200$ MHz, (ii) $\delta /2\pi =1$ GHz, (iii) $%
\gamma _{1\varphi ,fe}^{-1}=\gamma _{1\varphi ,fg}^{-1}=\gamma _{1\varphi
,eg}^{-1}=1$ $\mu $s, $\gamma _{1fg}^{-1}\sim \gamma _{1fe}^{-1}=10$ $\mu $%
s, $\gamma _{1eg}^{-1}=100$ $\mu $s [51,52], (iv) $\omega _{c1}/2\pi \sim 14$
GHz, $\omega _{c2}/2\pi \sim 9.0$ GHz, and $\omega _{c3}/2\pi \sim 1.0$ GHz,
(v)$\;\kappa _1^{-1}=\kappa _2^{-1}=\kappa _3^{-1}=10$ $\mu $s.

{\it Parameters for Step (ii)---}For the flux qutrit with three
levels shown in Fig.~1(b) [associated with the operation of step (ii)], we
choose $\omega _{eg}/2\pi \sim 1.0$ GHz and $\omega _{fe}/2\pi \sim 12.0$
GHz. Note that for a three-level flux qutrit, lowering the potential barrier
would slightly change the dipole coupling matrix element between the two
levels $\left| g\right\rangle $ and $\left| f\right\rangle $ or $\left|
e\right\rangle $ and $\left| f\right\rangle $ but significantly increase the
dipole coupling matrix element between the two lowest levels $\left|
g\right\rangle $ and $\left| e\right\rangle $. Thus, we can assume $\phi
_{fg}^{\prime }\sim \phi _{fg},\phi _{fe}^{\prime }\sim \phi _{fe},$ and $%
\phi _{eg}^{\prime }\sim 10\phi _{eg},$ where $\phi _{ij}^{\prime }$
represents the dipole coupling matrix element between the two levels $\left|
i\right\rangle $ and $\left| j\right\rangle $ for the level structure shown
in Fig.~1(b), with $ij\in \{fg,fe,eg\}.$ As a result, we have $%
g_{1eg}^{\prime }\sim g_{1fg}^{\prime }\sim g_{1fe}^{\prime }\sim g_1,$ $%
g_{2eg}^{\prime }\sim g_{2fg}^{\prime }\sim g_{2fe}^{\prime }\sim g_2,$ and $%
g_{3fg}^{\prime }\sim g_{3fe}^{\prime }\sim g_r.$ Other parameters used in
the numerical calculation for this step of operation are: (i) $\gamma
_{2\varphi ,fe}^{-1}=\gamma _{2\varphi ,fg}^{-1}=\gamma _{2\varphi
,eg}^{-1}=1$ $\mu $s, $\gamma _{2fg}^{-1}\sim \gamma _{2fe}^{-1}\sim $ $%
\gamma _{2eg}^{-1}=10$ $\mu $s, and (ii) the same values of $\omega
_{c1},\omega _{c2},\omega _{c3},\kappa _1^{-1},\kappa _2^{-1},\kappa _3^{-1}$
as those used for the operation of step (i).

\begin{figure}[tbp]
\begin{center}
\includegraphics[bb=0 0 575 370, width=12.5 cm, clip]{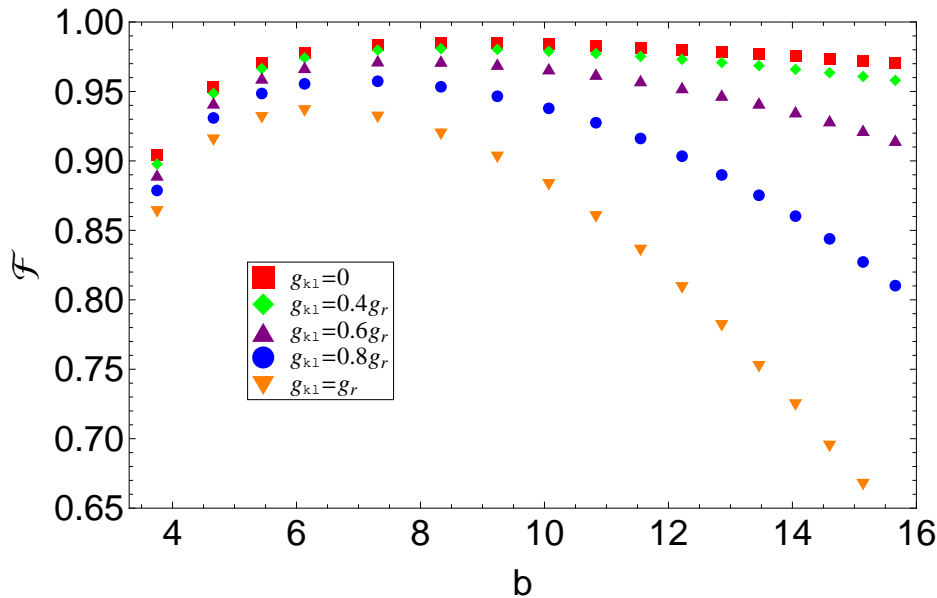} %
\vspace*{-0.08in}
\end{center}
\caption{(Color online) Fidelity versus $b$. Here, $b=\delta/g$. Refer to
the text for the parameters used in the numerical calculation. Here, $g_{kl}$
are the coupling strengths between cavities $k$ and $l$ ($kl=12,13,23$),
which are taken to be the same for simplicity.}
\label{fig:3}
\end{figure}

For the parameters chosen here, the fidelity versus $b\equiv \delta /g$ is
shown in Fig.~3, for $g_{kl}=0,0.4g_r,0.6g_r,0.8g_r,g_r$. Fig.~3 shows that
for $b\leq 12,$ when $g_{kl}\leq 0.4g_r$, the effect of intercavity cross
coupling between cavities on the fidelity of the prepared GHZ states is
negligible, which can be seen by comparing the top two lines. In addition,
one can see from Fig.~3 that for $b\sim 8,$ a high fidelity $\sim 98.5\%$
can be achieved when $g_{kl}\leq 0.4g_r$.

For $b\sim 8$, we have $g/\left( 2\pi \right) \sim $ $125$ MHz. In addition,
we set $g_r/2\pi =200$ MHz above. Note that a coupling constant $\sim 220$
MHz can be reached for a superconducting qubit coupled to a one-dimensional
CPW (coplanar waveguide) resonator [13]. For the cavity frequencies chosen
above and for the $\kappa _1^{-1},$ $\kappa _2^{-1}$ and $\kappa _3^{-1}$
used in the numerical calculation, the required quality factors for the
three cavities are $Q_1\sim 8.8\times 10^5,$ $Q_2\sim 5.7\times 10^5,$ and $%
Q_3\sim 6.3\times 10^4,$ respectively. Note that superconducting CPW
resonators with a loaded quality factor $Q\sim 10^6$ have been
experimentally demonstrated [53,54], and planar superconducting resonators
with internal quality factors $Q>10^6$ have also been reported recently
[55]. Our analysis given here demonstrates that preparation of the
three-cavity GHZ entangled photon Fock state is feasible within the present
circuit QED technique.

This condition, $g_{kl}\leq 0.4g_r,$ is not difficult to satisfy with
typical capacitive cavity-qutrit coupling illustrated in Fig.~2. As long as
the cavities are physically well separated, the intercavity cross-talk
coupling strength is $g_{kl}\sim g_r\left( C_k+C_l\right) /C_\Sigma $ (for $%
b\sim 8,$ $g$ is on the same order as $g_r$)$,$ where\textrm{\ }$C_\Sigma
=\sum_{j=1}^3C_j+C_q$ (see Fig. 2). For $C_j\sim 1$ fF and $C_\Sigma \sim
10^2$ fF (the typical values of the cavity-qutrit coupling capacitance and
the sum of all coupling capacitance and qutrit self-capacitance,
respectively), we have $g_{kl}\approx 0.02g_r$. Therefore, the condition $%
g_{kl}\leq 0.4g_r$ can be readily met in experiment. Hence, implementing
designs with sufficiently weak direct intercavity couplings is
straightforward. We remark that further investigation is needed for each
particular experimental setup. However, this requires a rather lengthy and
complex analysis, which is beyond the scope of this theoretical work.

\begin{center}
\textbf{V. CONCLUSION}
\end{center}

We have proposed a way for creating a three-cavity GHZ entangled photon Fock
state, by using a superconducting flux qutrit coupled to the cavities. This
proposal does not require the use of classical microwave pulses and any
measurement during the entanglement preparation. Our numerical simulation
shows that preparation of the three-cavity GHZ entangled photon Fock state
is feasible within the present circuit QED technique. We hope that this work
could stimulate further experimental and theoretical activities. Finally,
the method presented here is quite general, and can be applied to generate
the same type of entangled photon Fock state for triple cavities, which are
coupled by a different physical system with three levels, such as a
superconducting charge qutrit, a transmon qutrit, or a quantum dot.

Before we conclude, we should mention the previous work [56] which is
relevant to ours. Ref. [56] presents a scheme for the preparation of a
GHZ-type entangled photon coherent state of multiple cavities by having an
atom interact with each of the cavities dispersively and then measuring the
state of the atom. We are aware that a three-cavity GHZ entangled photon
Fock state can, in principle, be generated using the same procedure
described in Ref. [56]. However, the method in Ref. [56] requires a measurement
on the state of the atom; and since the prepared GHZ state depends on the
measurement outcome on the atomic states, the GHZ-state preparation is not
deterministic. In contrast, our present proposal mitigates these problems
effectively: there is no need to measure the state of the artificial atom,
and the generation of the GHZ state is deterministic. We believe that this
work is of interest because no measurement is needed and also no classical
microwave pulse is required, which greatly simplifies the operation and
reduces the engineering complexity and cost.

\begin{center}
\textbf{ACKNOWLEDGMENTS}
\end{center}

We thank Yang Yu for many fruitful discussions. This work was supported in
part by the National Natural Science Foundation of China under Grant No.
11074062, the Zhejiang Natural Science Foundation under Grant No. LZ13A040002,
the Open Fund from the SKLPS of ECNU, and the funds from Hangzhou
Normal University under Grant No. HSQK0081. Q.P. Su was supported by Zhejiang
Provincial Natural Science Foundation of China (Grant No. LQ12A05004).
Z. F. Zheng was supported by Student Scientific Research Foundation of Hangzhou
Normal University under Grant No. 1283XXM101.


\begin{references}
\bibitem{s50} For the operation of step (ii), we assumed $g_{3fg}^{\prime }\sim
g_{3fe}^{\prime }\sim g_r$ due to $\phi _{fg}^{\prime }\sim \phi
_{fe}^{\prime }\sim $ $\phi _{eg}^{\prime }.$ In addition, because of $\phi
_{fg}^{\prime }\sim \phi _{fg}, \phi _{fe}^{\prime }\sim \phi _{fe},$ and $%
\phi _{eg}^{\prime }\sim 10\phi _{eg},$ we have $g_{3fg}^{\prime }\sim
g_{3fg},$ $g_{3fe}^{\prime }\sim g_{3fe},$ and $g_r\sim 10g_{3eg}.$ Combining
these, we have $g_{3eg}\sim 0.1g_r,$ $g_{3fe}\sim g_{3fg}\sim g_r.$
\bibitem{s51} T1 (energy relaxation time) and T2 (dephasing time) can be made to be a few tens of $\mu $s for
the state-of-art superconducting devices (e.g. see Ref. 18).
\bibitem{s52} For a flux qutrit with the three levels considered, the $\left|
g\right\rangle \leftrightarrow \left| e\right\rangle $ dipole matrix
element is much smaller than that of the $\left| g\right\rangle \leftrightarrow
\left| f\right\rangle $ and  $\left|
e\right\rangle \leftrightarrow \left| f\right\rangle $ transitions. Thus, $\gamma _{eg}^{-1}\gg \gamma _{fg}^{-1},\gamma _{fe}^{-1}.$
\end{references}
\end{document}